\documentclass[twocolumn,trackchanges]{aastex631}

\usepackage{color}
\shorttitle{Climates of oblique planets}
\shortauthors{Kodama et al.}
\graphicspath{{./}{figures/}}

\begin{document}

\title{Climate of high obliquity exo-terrestrial planets with a three-dimensional cloud system resolving climate model}

\correspondingauthor{Takanori Kodama}
\email{koda@g.ecc.u-tokyo.ac.jp}

\author[0000-0001-9032-5826]{Takanori Kodama} 
\affiliation{Komaba Institute for Science, The University of Tokyo, 3-8-1 
Komaba, Meguro, Tokyo 153-8902, Japan}

\author[0000-0002-9321-6403]{Daisuke Takasuka} 
\altaffiliation{co-first author}
\affiliation{Japan Agency for Marine-Earth Science and Technology, 3173-25
Shouwa, Kanazawa, Yokohama, Kanagawa 236-0001, Japan}

\author[0000-0002-4378-9166]{Sam Sherriff-Tadano} 
\affiliation{School of Earth and Enviroment, University of Leeds, Woodhouse Lane, Leeds, LS2 9JT, UK}

\author[0000-0003-4789-4737]{Takeshi Kuroda} 
\affiliation{Department of Geophysics, Graduate School of Science,
Tohoku University, 6-3 Aramaki-aza-Aoba, Aoba, Sendai, Miyagi 980-8578, Japan}
\affiliation{Division for the Estanblishment of Frontier Science, Organization for Advanced Studies, Tohoku University, 2-1-1 Katahira, Aoba,Sendai Miyagi 980-8577, Japan}

\author[0000-0001-8465-1967]{Tomoki Miyakawa} 
\affiliation{Atmosphere and Ocean Research Institute,
The University of Tokyo, 5-1-5, Kashiwanoha, Kashiwa, Chiba 277-8568, Japan}

\author[0000-0003-1745-5952]{Ayako Abe-Ouchi} 
\affiliation{Atmosphere and Ocean Research Institute,
The University of Tokyo, 5-1-5, Kashiwanoha, Kashiwa, Chiba 277-8568, Japan}
\affiliation{Japan Agency for Marine-Earth Science and Technology, 3173-25
Shouwa,Kanazawa, Yokohama, Kanagawa 236-0001, Japan}

\author[0000-0003-3580-8897]{Masaki Satoh} 
\affiliation{Atmosphere and Ocean Research Institute,
The University of Tokyo, 5-1-5, Kashiwanoha, Kashiwa, Chiba 277-8568, Japan}



\begin{abstract}

Planetary climates are strongly affected by planetary orbital parameters such as obliquity, eccentricity, and precession. In exoplanetary systems, exo-terrestrial planets should have various obliquities. High-obliquity planets would have extreme seasonal cycles due to the seasonal change of the distribution of the insolation. Here, we introduce the Non-hydrostatic ICosahedral Atmospheric Model (NICAM), a global cloud-resolving model, to investigate the climate of high-obliquity planets. This model can explicitly simulate a three-dimensional cloud distribution and vertical transports of water vapor. We simulated exo-terrestrial climates with high resolution using the supercomputer FUGAKU. We assumed aqua-planet configurations with 1 bar of air as a background atmosphere, with four different obliquities ($0^{\circ}$, $23.5^{\circ}$, $45^{\circ}$, and $60^{\circ}$). We ran two sets of simulations: 1) low-resolution ($\sim 220$ km-mesh as the standard resolution of a general circulation model for exoplanetary science) with parametrization for cloud formation, and 2) high-resolution ($\sim 14$ km-mesh) with an explicit cloud microphysics scheme. Results suggest that high-resolution simulations with an explicit treatment of cloud microphysics reveal warmer climates due to less low cloud fraction and a large amount of water vapor in the atmosphere. It implies that treatments of cloud-related processes lead to a difference between different resolutions in climatic regimes in cases with high obliquities.

\end{abstract}



\section{Introduction} \label{sec:intro}

Planetary obliquity is a key astronomical parameter that strongly affects planetary climate by controlling the distribution of incoming stellar radiation at the top of the atmosphere. The Earth's obliquity is relatively stable at around $23.5^{\circ}$ but it has varied from $22^{\circ}$ to $24.5^{\circ}$ with a period of $41,000$ yr. During the Earth's history, glaciation-deglaciation cycles may have been controlled by the changes in the distribution of insolation due to changes in obliquity, and this is called Milankovitch orbital insolation forcing \citep{Abe-Ouchi+2013}. Also, high-obliquity planets should have extreme seasonal cycles due to the seasonal change of the distribution of the insolation. When the obliquity exceeds $54^{\circ}$, the annual average insolation at the poles is larger than that at the equator. In our solar system, planets have various obliquities because of the tidal influence from the Sun and past collisions. From theoretical prediction, Mars' obliquity may have varied between $0^{\circ}$ and $60^{\circ}$ in its history \citep{Laskar&Robutel1993}. For the Earth, without the Moon, its obliquity would vary due to the solar tides between $0^{\circ}$ and $90^{\circ}$ over a timescale of less than $10$ Myr \citep{Laskar+1993}.

Since 1995, we found over $5,000$ exoplanets, including candidates. Some of them are thought to be Earth-like rocky planets within the habitable zone of their host stars, where liquid water remains stable on their surface. In the next decade, such potentially habitable planets -- for example, Proxima Centauri b and Trappist-1 planets -- will be primary targets for observation of biosignatures. Classically, the edges of the habitable zone have been investigated with aqua planet configuration (water-covered surface) using a one-dimensional radiative-convective model \citep[e.g.][]{Abe&Matsui1988,Kasting+1993, Nakajima+1992, Kopparapu+2013}. Recently, it is possible to estimate climates for potential habitable exo-terrestrial planets using three-dimensional general circulation models (GCMs)\citep[e.g.][]{Ishiwatari+2002, Abe+2011, Leconte+2013, Wolf&Toon2015, Turbet+2016, Way+2018, Turbet+2018, Kodama+2018, Kodama+2019, Kodama+2021}. Most of the detected potentially habitable planets are in systems of M-type stars because it is relatively easier to detect terrestrial planets around M-type stars compared to around G-type stars, like the Sun, with current exoplanet observation capabilities. Such terrestrial planets within the habitable zone of M-type stars are thought to be in a tidally locked state. Tidally locked exoplanets should have an obliquity of $0^{\circ}$ with a rotation period equal to the orbital period, and a permanent day-side and night-side. GCM studies showed that cloud formation on their day-side is important for maintaining their habitability \citep{Yang+2013, Kopparapu+2016, Kopparapu+2017, Haqq-Misra+2018}.

Terrestrial planets within the habitable zone of a G-type star will also be good targets in the near future for the next flagship space telescope. Terrestrial exoplanets around G-type stars should have various obliquities and it is crucial to understand the climates of high-obliquity terrestrial planets. Traditionally, the climates of high-obliquity planets have been estimated using a one-dimensional zonally averaged energy-balance climate model (EBCM). \cite{Williams&Kasting1997} estimated the effect of high obliquity on the Earth's climate using an EBCM, and found that the summer time surface temperatures over middle and high latitudes became $50^{\circ}\mathrm{C} - 80^{\circ}\mathrm{C}$ as the planetary obliquity approaches $90^{\circ}$. \cite{Armstrong+2014} investigated the outer edge of the Habitable zone using EBCM considering the effect of orbital parameters. Idealized GCM studies also showed high surface temperatures over the continent at middle and high latitudes during summer on high-obliquity planets. \cite{william&pollard2003} investigated the climates of Earth-like planets with different obliquities from $0^{\circ}$ to $85^{\circ}$ using an idealized GCM, GENESIS 2, and showed a surface temperature of $80^{\circ}\mathrm{C}$ - $100^{\circ}\mathrm{C}$ at the middle and high latitudes of large continents for an obliquity larger than $54^{\circ}$. Recent studies on high-obliquity planets using GCMs showed that wide seasonally reversing Hadley cells transport energy from the summer hemisphere to the winter hemisphere, and dominate the planet's hydrological cycles \citep{Lobo&Bordoni2020}. \cite{Kang2019} also investigated the climate of such a high-obliquity planet and the roles of ice-albedo feedback and cloud radiation feedback, which reduce the absorbed solar radiation, and pointed out the important role of clouds in these feedback mechanisms.The effect of orbital parameters in the outer edge of the habitable zone is also investigated with GCM \citep{Linsenmeier2015}. Additionally, the climates of high-obliquity planets can be addressed using a coupled ocean-atmosphere model. For a high-obliquity planet, a coupled ocean-atmosphere model has shown the role of ocean dynamics on global climate, and that an ocean remains warmer at high latitudes than at the equator \citep{Ferreira+2014}. Stable climates for planets with high obliquities are investigated using EBCMs and GCMs, and the hysteresis structure of possible climates is discussed \citep{Kilic+2017, Kilic+2018, Rose+2017}. For terrestrial exoplanets around M dwarf, the relation between their climate and obliquity has been investigated using GCM. \cite{Wang+2016} found that such planet with high obliquity has less low cloud fraction and gets warmer than planets with $0^\circ$ obliquity.

Clouds pose large uncertainties in models representing the climate of potentially habitable exoplanets, as well as Earth models. Traditionally, conventional GCMs with a $O(10^2)$-km horizontal mesh have used cumulus parameterizations and large scale condensation schemes to evaluate cloud-related processes that cannot be explicitly resolved in model grids. Cumulus parameterizations estimate the changes in temperature and moisture and precipitation in $O(10^2)$-km scales by evaluating vertical transports of heat and moisture as a result of an ensemble of individual convective processes such as condensation, evaporation, and turbulent motions \cite[e.g.][]{Manabe+1965, Arakawa&Schubert1974, Kuo1974, Betts&Miller1986}. Similarly, large-scale condensation schemes represent the condensation processes of clouds except for cumulus convection, which affects temperature and moisture budgets associated with stratiform clouds and radiative fluxes through cloud fraction \citep{Le_Trent&Li1991}.

As one powerful approach to reduce the uncertainties associated with cloud processes in a climate model, three-dimensional high-resolution global models (with less than about $10$-km horizontal mesh) have been developed to resolve cumulus cloud systems explicitly; such a model is called a global cloud/cloud-system-resolving model (GCRM) \citep[e.g.,][]{Satoh+2019, Stevens+2019}. GCRMs are actively used to investigate the mechanisms of Earth's weather such as the Madden--Julian oscillation \citep[e.g.,][]{Miura+2007, Miyakawa+2014} and tropical cyclones \citep[e.g.,][]{Yamada&Satoh2013}, and increase our understanding of multi-scale convective systems. In recent years, growing computational power allows GCRMs with high resolutions to be used, together with explicit treatment of cloud microphysics for the long periods ($O(10)$ years), as indicated by the AMIP-type climate simulations \citep{CKodama+2015,CKodama+2021}. Although ``the explicit treatment of cloud microphysics" here simply means that the mixing ratio of water substances is calculated by many bulk equations that directly represent the cloud microphysics processes (e.g., evaporation, melting, droplet collection) without cumulus convection schemes and it does not necessarily mean the representation of individual cloud particles, this provides much more information on the interaction among clouds, water vapor, and dynamics, and thus leads to a more accurate assessment of the impact of clouds on the climate. 

Although we do not know whether the explicit treatment of convection without cumulus parameterizations is required for better reproduction the exoplanet climate, it is important to understand how GCRMs represent the equilibrium states under boundary conditions far from those on the Earth. GCRMs are free from umbiguity related to the details of cumulus parameterization schemes, and the results from GCRMs are more physically approachable. Thus, as our first step, we address the climates of high-obliquity planets simulated using GCRMs, comparing climates with a setting corresponding to conventional GCMs. In particular, we mainly focus on the climatological states of temperature, precipitation, and large-scale circulation, which are directly related to the interaction among clouds, convection, and moisture fields.

In section 2, we describe the model and numerical experimental setup used in this study. In section 3, we show the climatological mean states simulated with various obliquity values in cases for a low-resolution run with a cumulus parameterization scheme (a GCM run) and a high-resolution run with explicit treatment of convection (a GCRM run). In section 4, we discuss the effect of the amount of $\mathrm{CO}_{2}$, and the size and location of a continent on the climate for a high-obliquity planet, and the climates for potentially habitable exo-terrestrial planets. In section 5, we summarize our findings.

\section{Methods} \label{sec:methods}

\subsection{Model}
In this study, we used the Nonhydrostatic Icosahedral Atmospheric Model (NICAM) \citep{Tomita&Satoh2004, Satoh+2008, Satoh+2014}, which is a GCRM. NICAM is a finite-volume global model that solves the non-hydrostatic Euler equations using a geodesic grid \citep{Tomita+2002}. To increase the horizontal resolution above $\sim 10$ km globally, the governing equations must be non-hydrostatic. NICAM uses a terrain-following vertical  coordinate system with Lorenz staggering \citep{Tomita&Satoh2004}. For radiative transfer, MstrnX with a correkated-k method \citep{Sekiguchi&Nakajima2008} is used. The level 2 of the Mellor-Yamada-Nakanishi-Niino (MYNN) scheme is used to parameterize turbulent fluxes in the boundary layer \citep{Mellor&Yamada1974, Nakanishi&Niino2009}. As described in the next subsection, the treatment of cumulus convection and precipitation processes depends on the horizontal resolution.


\subsection{Setting}
We investigated the climate for planets with various obliquities. We adopted aquaplanet configurations with four different obliquities ($\phi = 0^{\circ}, 23.5^{\circ}, 45^{\circ}$, and $60^{\circ}$). In this study, we ran two sets of simulations; one is a low-resolution simulation ($\sim 220$-km horizontal mesh) with a convective parameterization scheme (Chikira-Sugiyama scheme, \citet{Chikira&Sugiyama2010}) and a large-scale condensation scheme for cloud formation \citep{Le_Trent&Li1991}, and the other is a high-resolution simulation ($\sim 14$-km horizontal mesh) with a cloud microphysics scheme (NSW6) alone. The single-moment bulk cloud microphysics scheme with six water categories (NSW6) \citep{Tomita2008} calculates the mixing ratio of their categories explicitly, instead of cumulus convective parameterizations. While our low-resolution simulations correspond to conventional GCM studies for current exoplanetary science, our high-resolution simulations can be viewer as the highest resolution simulations conducted for a global climate study in exoplanetary science. For the vertical resolution, we used $40$ vertical layers with a model top of about $40$ km. 

We started all the experiments from an isothermal atmospheric state with $300$ K and SST (sea surface temperature) distributions obtained from the ``Qobs" setting \citep{Neale&Hoskins2000}, which is zonally symmetric and has a meridional gradient mimicking that on the Earth with a temperature of $300$ K at the equator. After the initialization of simulations, lower boundary conditions are predicted by a global slab ocean model having a mixed layer depth of $50$ m, corresponding to a typical value on the Earth. The sea surface with its temperature below $271.15$ K set to be sea ice. As a background atmosphere, we assumed an Earth-like planet with $1$ bar of air containing $348$ ppm of $\mathrm{CO}_{2}$ and meridionally symmetric ozone distributions made by annual and annular means of realistic distributions of the Earth. The rotation velocity, orbital period, and planetary mass and radius were set to be the same as those for the Earth. The eccentricity was set to zero to focus on the effect of the obliquity on the climate. The time steps were $20$ min and $1$ min for the low-resolution and high-resolution runs, respectively. Low- and high-resolution simulations were integrated for $15$ yrs. We analyzed the climate for the last $5$ yr. This integration period in our simulation is a bit short to reach the equilibrium state, but its state is almost reached in all cases.

\section{Climates of oblique planets} \label{sec:climate}

\subsection{Low resolution with parameterization for clouds}
\begin{figure*}[htbp!]
\begin{center}
\includegraphics[scale=0.65]{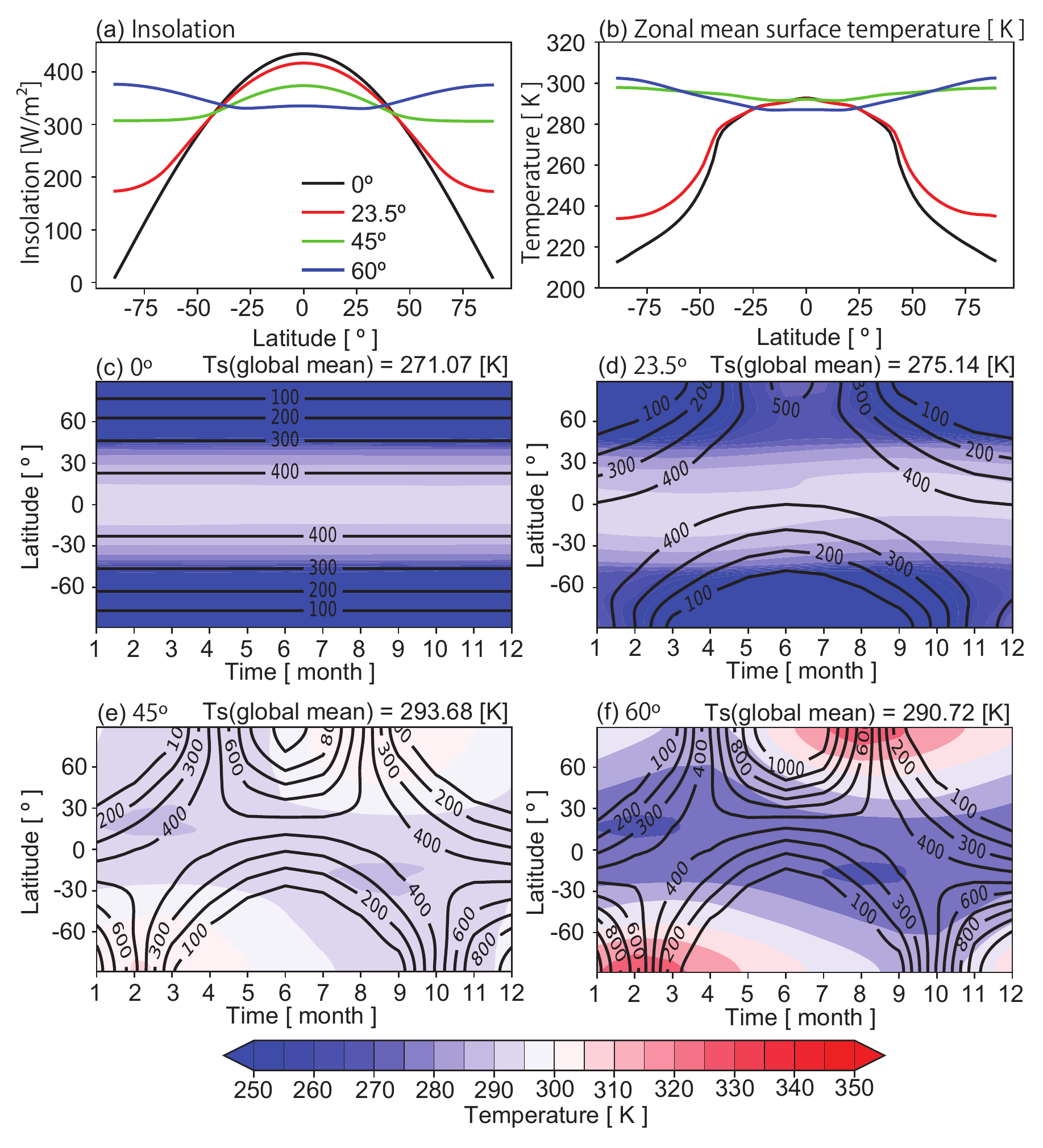}
\caption{Annual mean zonally averaged insolation distribution and surface temperature are shown in (a) and (b). Climatological seasonal march of zonal mean surface temperature in color (K) and the insolation in contours (W/$\mathrm{m}^{2}$) for (c) $\phi=0^\circ$, (d) $\phi=23.5^\circ$, (e) $\phi=45^\circ$, and (f) $\phi=60^\circ$. The global mean surface temperatures are shown in the upper right on each panel. All cases were run with a low resolution and parameterization for clouds.}
\label{obl_fig1}
\end{center}
\end{figure*}

First, we show the climate for planets with various obliquities in cases for a low-resolution run with parameterization for clouds (a GCM run). Figures \ref{obl_fig1} (a) and (b) show the annual mean zonally averaged insolation and surface temperature distribution, respectively. In Figure \ref{obl_fig1}(b), the merdional gradient of the surface temperature is opposite between the low ($\phi = 0^{\circ}$, $23.5^{\circ}$) and high obliquity cases ($\phi = 45^{\circ}$, $60^{\circ}$). Compared to Figure \ref{obl_fig1}(a), this temperature distributions follow the insolation pattern except for $\phi=45^\circ$. A possible reason for the mismatch between the annual-mean solar insolation and surface temperature meridional gradient for $\phi=45^\circ$ is shortwave cooling due to abundant low clouds in the tropics (see Figs. \ref{obl_fig7}e and \ref{obl_fig8}e). 

Figures \ref{obl_fig1} (c)-(f) show the seasonal change of the zonally-averaged surface temperature and solar insolation. For low obliquity cases, larger insolation is observed at low latitudes than at middle and high latitudes. Following this insolation, the surface temperature is warmer at low latitudes than at middle-to-high latitudes. Meanwhile, for high obliquity cases, solar insolation and surface air temperature are larger and higher, respectively, at middle-to-high latitudes than at low latitudes, which results in a temperature gradient opposite to that for low-obliquity cases. These results are in good agreement with a previous study for high-obliquity planets, despite the difference of used mixed layer depth  \citep{Lobo&Bordoni2020}.

\begin{figure*}[htbp!]
\begin{center}
\includegraphics[scale=0.65]{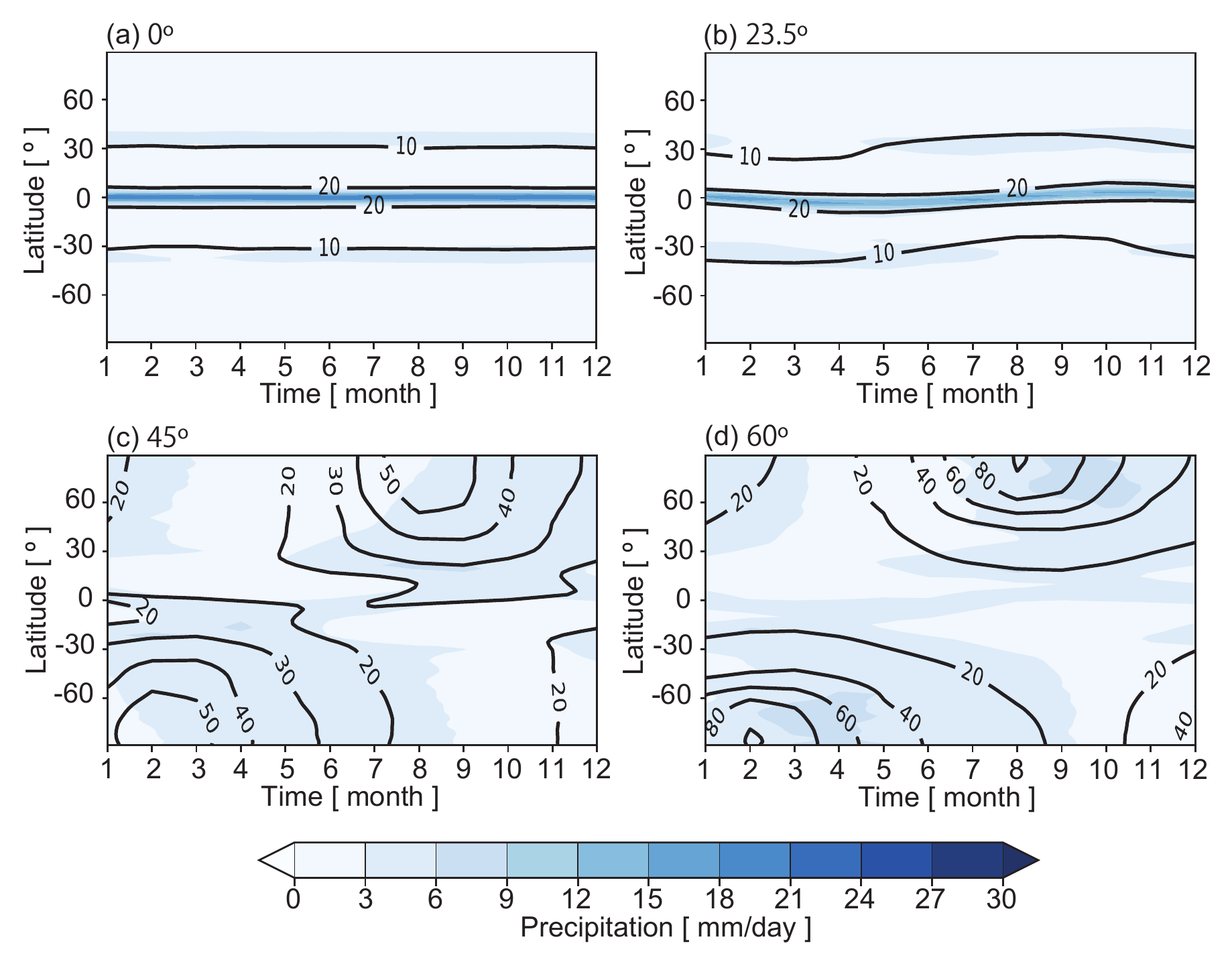}
\caption{Climatological seasonal march of zonal mean precipitation in color (mm/day) and the column water vapor in contours (kg/$\mathrm{m}^{2}$) for (a) $\phi=0^\circ$, (b) $\phi=23.5^\circ$, (c) $\phi=45^\circ$, and (d) $\phi=60^\circ$. All cases were run with a low resolution and parameterization for clouds.}
\label{obl_fig2}
\end{center}
\end{figure*}

Figure \ref{obl_fig2} shows the seasonal change of the zonally averaged precipitation and column water vapor content. Low-obliquity planets have heavier precipitation at low latitudes than at middle to high latitudes. As the obliquity increases, precipitation and moisture distributions drastically change; high obliquity planets have relatively strong precipitation at middle to high latitudes from the end of the summer to the winter in comparison with the tropics. These distributions are consistent with the fact that more humid environments are observed at middle to high latitudes in summer.

\begin{figure*}[ht!]
\begin{center}
\includegraphics[scale=0.65]{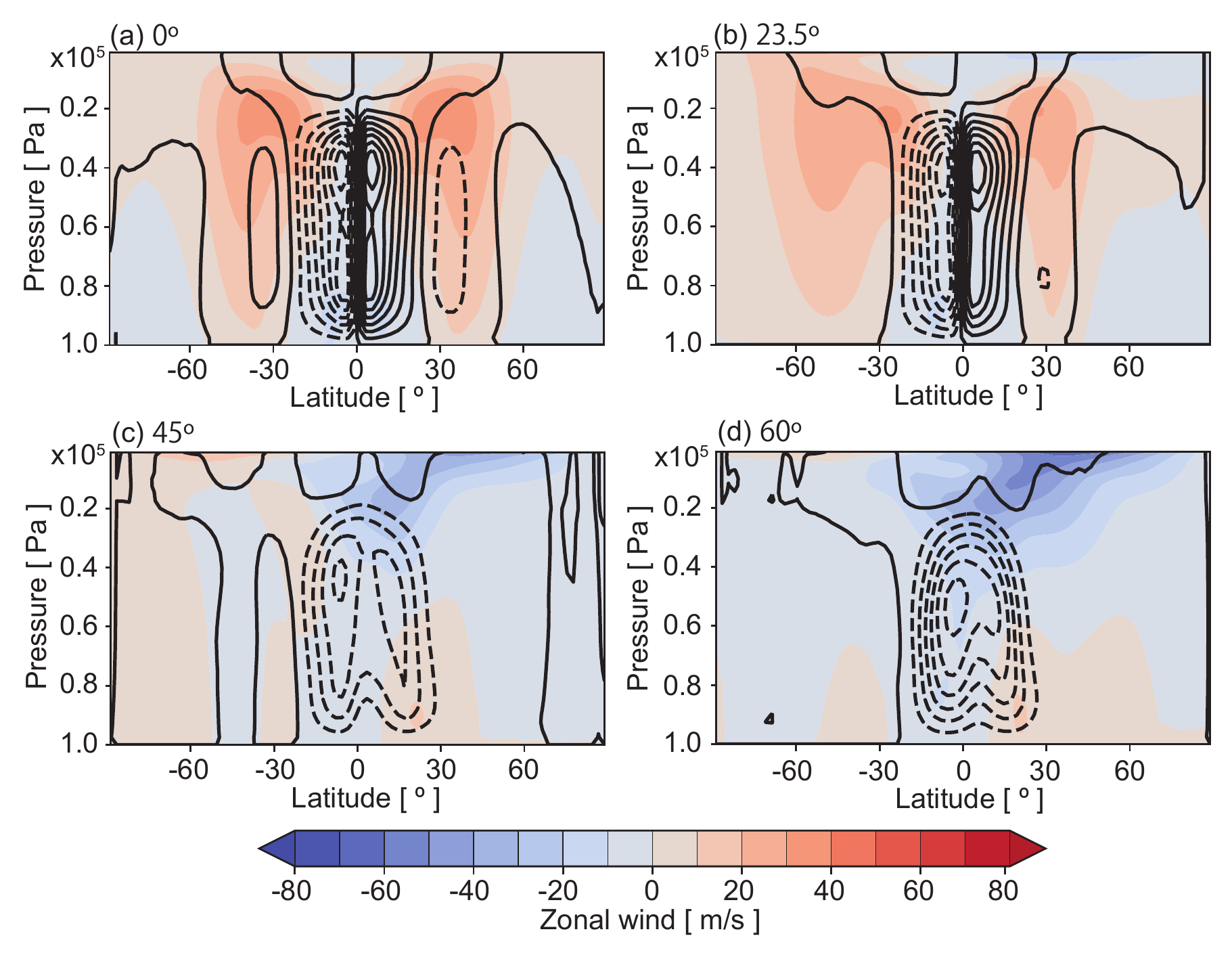}
\caption{Vertical cross sections of zonal mean zonal wind (color; m/s) and the mass stream function (contours; kg/s)} for (a) $\phi=0^\circ$, (b) $\phi=23.5^\circ$, (c) $\phi=45^\circ$, and (d) $\phi=60^\circ$, averaged in boreal summer (June-July-August). All cases were run with a low resolution and parameterization for clouds. The contour intervals were set to $5 \times 10^{10}$ kg/s. 
\label{obl_fig3}
\end{center}
\end{figure*}

Figure \ref{obl_fig3} shows the mass stream function and the zonal mean zonal winds averaged in boreal summer (June-July-August). For high-obliquity cases, meridional circulations composed of only one cell are dominant. Also, they have an ascent region around the equator, which may correspond to the ITCZ (inter-tropical convergence zone)-like precipitation (Fig. \ref{obl_fig2}). Note that two-cell meridional circulations for $\phi=23.5^\circ$, which are different from those in JJA on the present Earth, possibly arise from the weaker meridional temperature gradient due to our aqua-planet configurations with 50-m mixed layer depth.

\subsection{High resolution with cloud microphysics}
\begin{figure*}[htbp!]
\begin{center}
\includegraphics[scale=0.65]{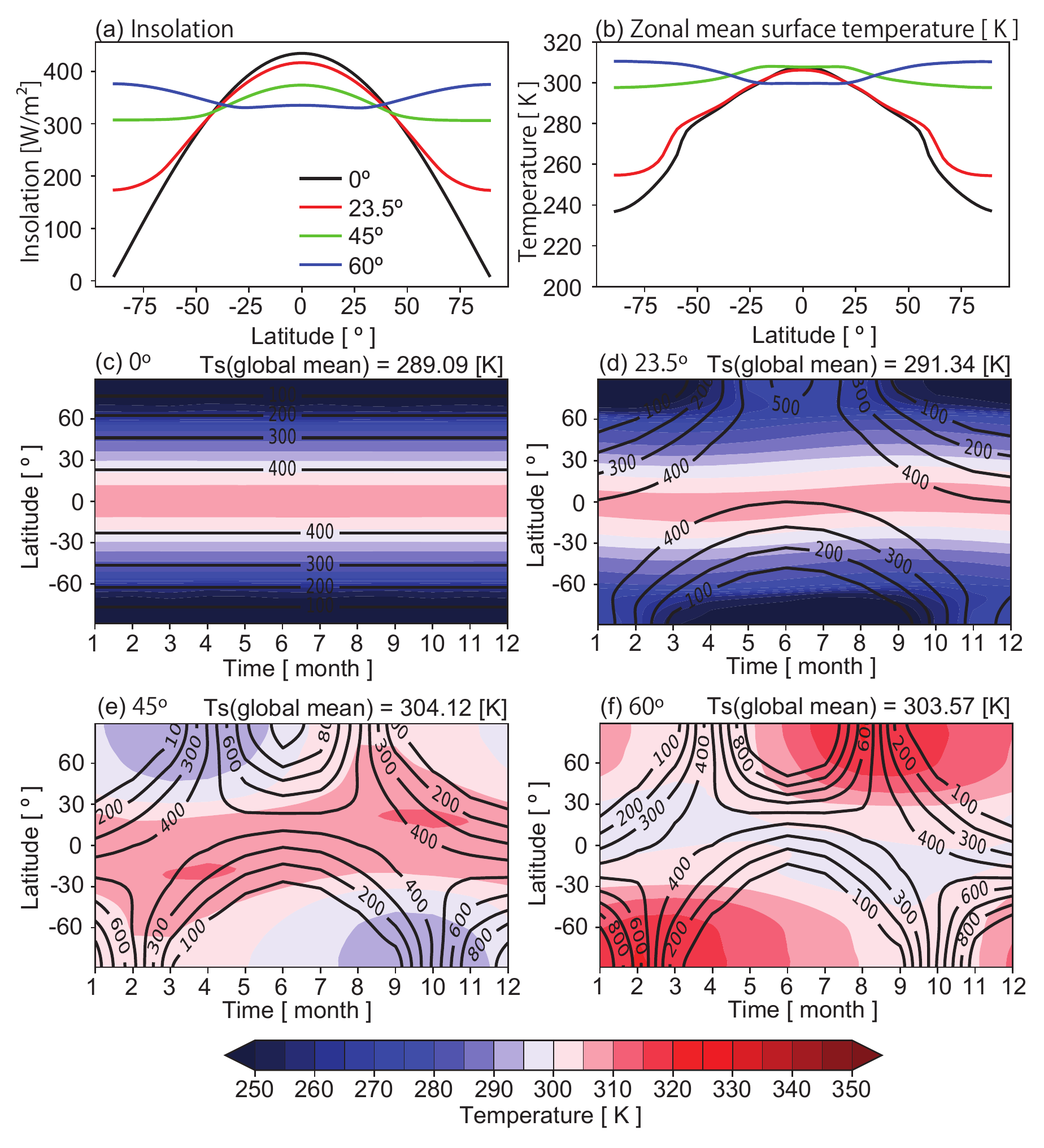}
\caption{The same as Figure \ref{obl_fig1}, except for high-resolution runs with explicit treatment of cloud microphysics.}
\label{obl_fig4}
\end{center}
\end{figure*}

Next, we present the results for a high-resolution run with explicit treatment of convection without parameterization for clouds (a GCRM run). As for the surface temperature and its relationship with solar insolation, there are some noteworthy differences between high-resolution (Fig. \ref{obl_fig4}) and low-resolution runs (cf. Fig. \ref{obl_fig1}). One is that the surface temperature is much higher in high-resolution runs than in low-resolution runs for all obliquities. Another is that the obliquity for which the meridional gradient of surface temperature is reversed is changed; in high-resolution runs, the poleward upgradient temperature gradient, which is confirmed for $\phi = 45^\circ$ and $60^\circ$ cases in low-resolution runs (Fig. \ref{obl_fig1}b), is realized only for the $\phi = 60^\circ$ case (Fig. \ref{obl_fig4}b). 

\begin{figure*}[htbp!]
\begin{center}
\includegraphics[scale=0.65]{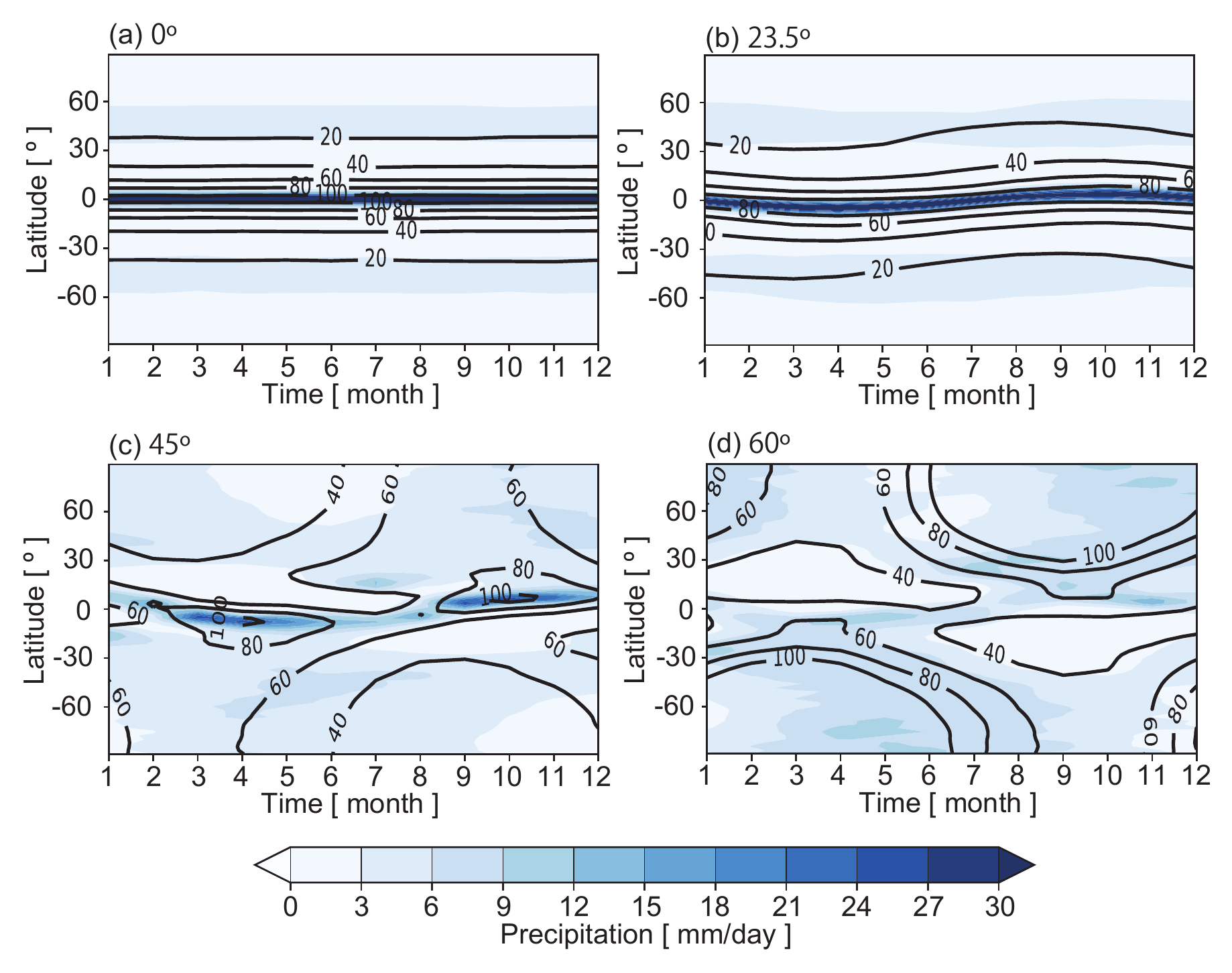}
\caption{The same as Figure \ref{obl_fig2}, except for high-resolution runs with explicit treatment of cloud microphysics.}
\label{obl_fig5}
\end{center}
\end{figure*}

In Figure \ref{obl_fig5}, which shows precipitation and column-integrated water vapor distributions for high-resolution runs, we find more precipitation and more humid climates for all obliquities than in the low-resolution runs. The reasons why the high-resolution run has the more humid environment are speculated in terms of both convection characteristics and large-scale mean states. As for the former, the explicit cloud scheme cannot generate deep clouds until the atmospheric column reaches almost saturated states, which results in the longer time scale of convective adjustment and more moisture accumulation without releasing the convective instability instantly than for the low-resolution run. In addition to this, higher surface temperature, which is related to less low cloud fraction (see Figure \ref{obl_fig8}), can also contribute to more water vapor following the Clausius-Clapeyron relation. 

The precipitation patterns for low obliquities are almost the same as that in the low-resolution runs; they have a precipitation band around low latitudes. Note that this ITCZ structure is realized even in a high obliquity case such as $\phi = 60^{\circ}$ for both high- and low-resolution runs. Also, as in the  low-resolution run (Fig. \ref{obl_fig2}d), the precipitation in high-obliquity planets is realized from the end of the summer to the winter in the high-resolution run (Fig. \ref{obl_fig5}d).

\begin{figure*}[htbp!]
\begin{center}
\includegraphics[scale=0.65]{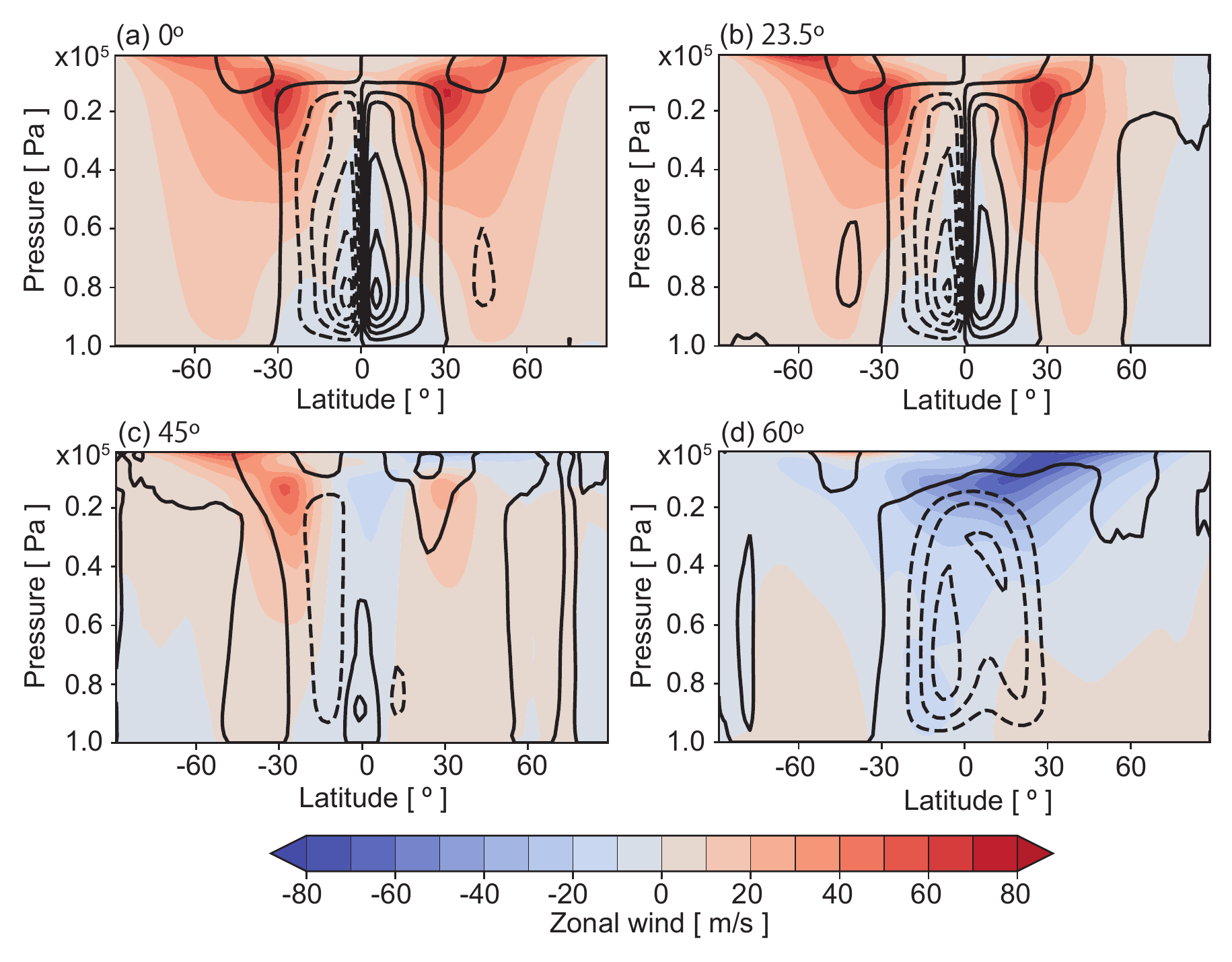}
\caption{The same as Figure \ref{obl_fig3}, except for high-resolution runs with explicit treatment of cloud microphysics.}
\label{obl_fig6}
\end{center}
\end{figure*}

As with Figure \ref{obl_fig3}, Figure \ref{obl_fig6} shows the mass stream function and the zonal mean zonal winds averaged in boreal summer for the high-resolution runs. For cases with $\phi \leq 45^{\circ}$, direct circulation composed of two cells having the ascent branch is established in the tropics and subtropics, although its strength is much weaker for $\phi = 45^{\circ}$ than for the other two obliquity cases. For $\phi = 60^{\circ}$, a one-cell circulation is obvious and it has ascent regions around 30$^\circ$N and at the equator.

\subsection{Low resolution versus high resolution}
A comparison of high- and low-resolution cases shows some climatic features different from each other for the same obliquities (i.e., the same insolation distributions). For brief comparisons, Table \ref{tab:summary} summarizes global- and annual-mean climatic variables, including the surface temperature, precipitation, water vapor amount, low- and high-level cloud fractions, the cloud radiative forcing. 

For all obliquities, high-resolution runs with the explicit treatment of microphysics have more precipitation and water vapor and higher surface temperature than low-resolution runs for our model settings. Furthermore, a different climatic regime is realized for the obliquity $\phi=45^{\circ}$ between high- and low-resolution runs; while the meridional surface temperature gradient is downward in the poleward direction for high-resolution cases, the opposite gradient is realized for low-resolution cases. Related to this, high-resolution runs have obvious ITCZ-like precipitation bands around the equator, whereas precipitation for low-resolution runs tends to have a peak at middle to high latitudes.

\begin{figure*}[htbp!]
\begin{center}
\includegraphics[scale=0.65]{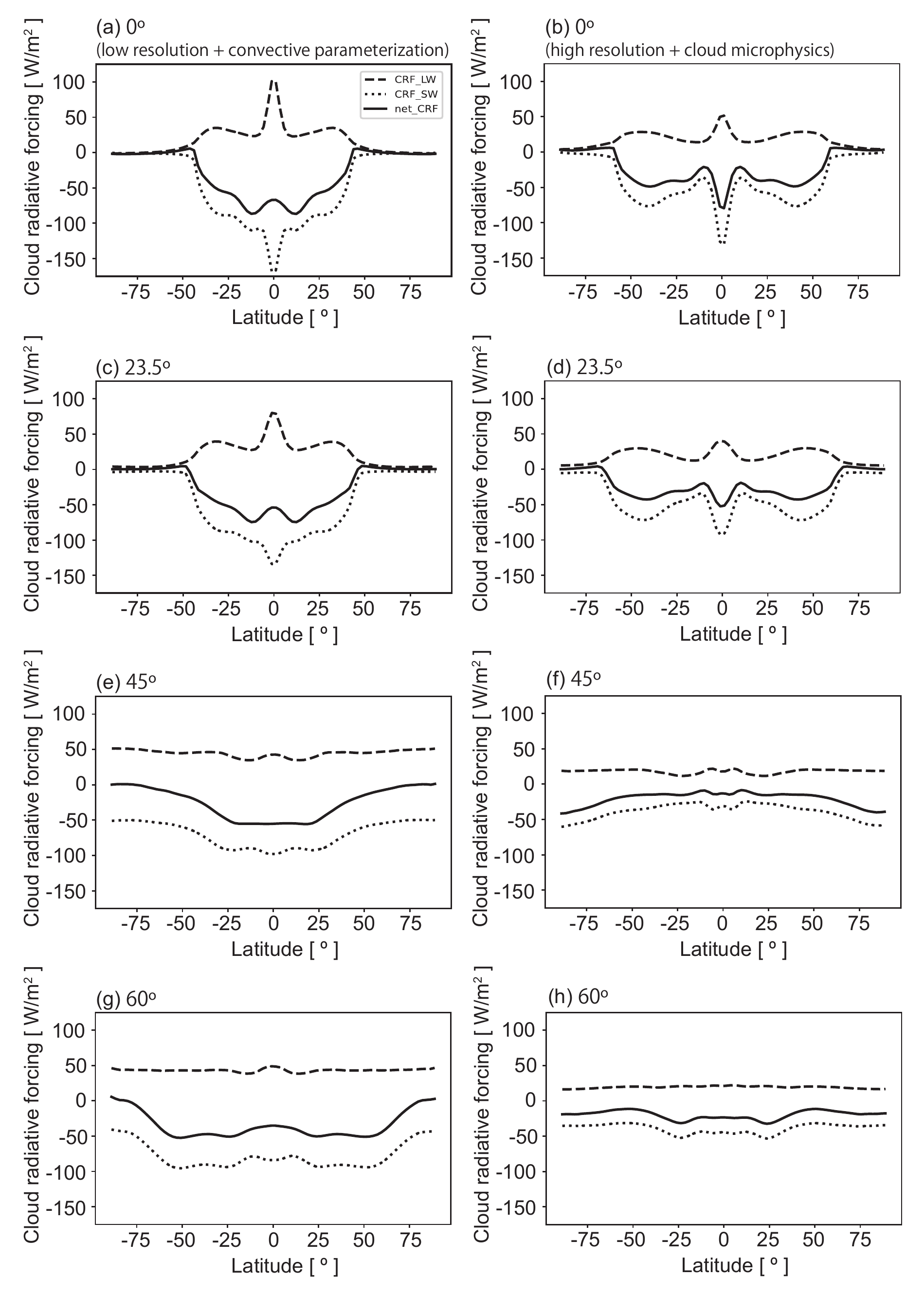}
\caption{Zonally averaged cloud radiative forcing at low resolution with convective parameterization (left column) and at high resolution with explicit treatment of cloud microphysics (NSW6) (right column). The net cloud radiative forcing, the cloud shortwave radiative forcing and the cloud longwave radiative forcing are represented in the solid, dotted and dashed lines, respectively.}
\label{obl_fig7}
\end{center}
\end{figure*}

\begin{figure*}[htbp!]
\begin{center}
\includegraphics[scale=0.65]{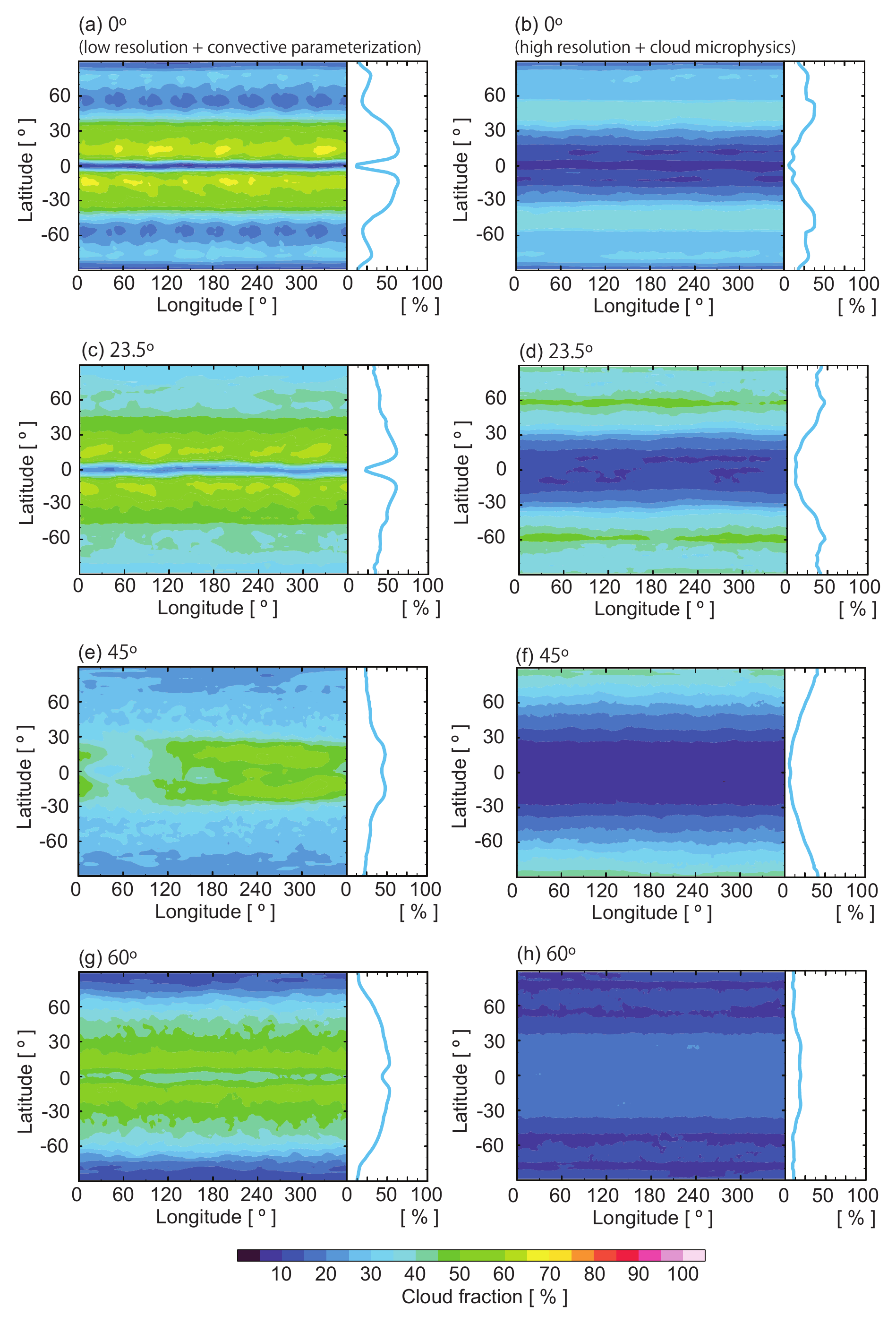}
\caption{Maps for cloud fraction of low clouds at low resolution with convective parameterization (left column) and at high resolution with explicit treatment of cloud microphysics (NSW6) (right column). Right panels beside maps show zonally averaged cloud fraction.}
\label{obl_fig8}
\end{center}
\end{figure*}

\begin{figure*}[htbp!]
\begin{center}
\includegraphics[scale=0.65]{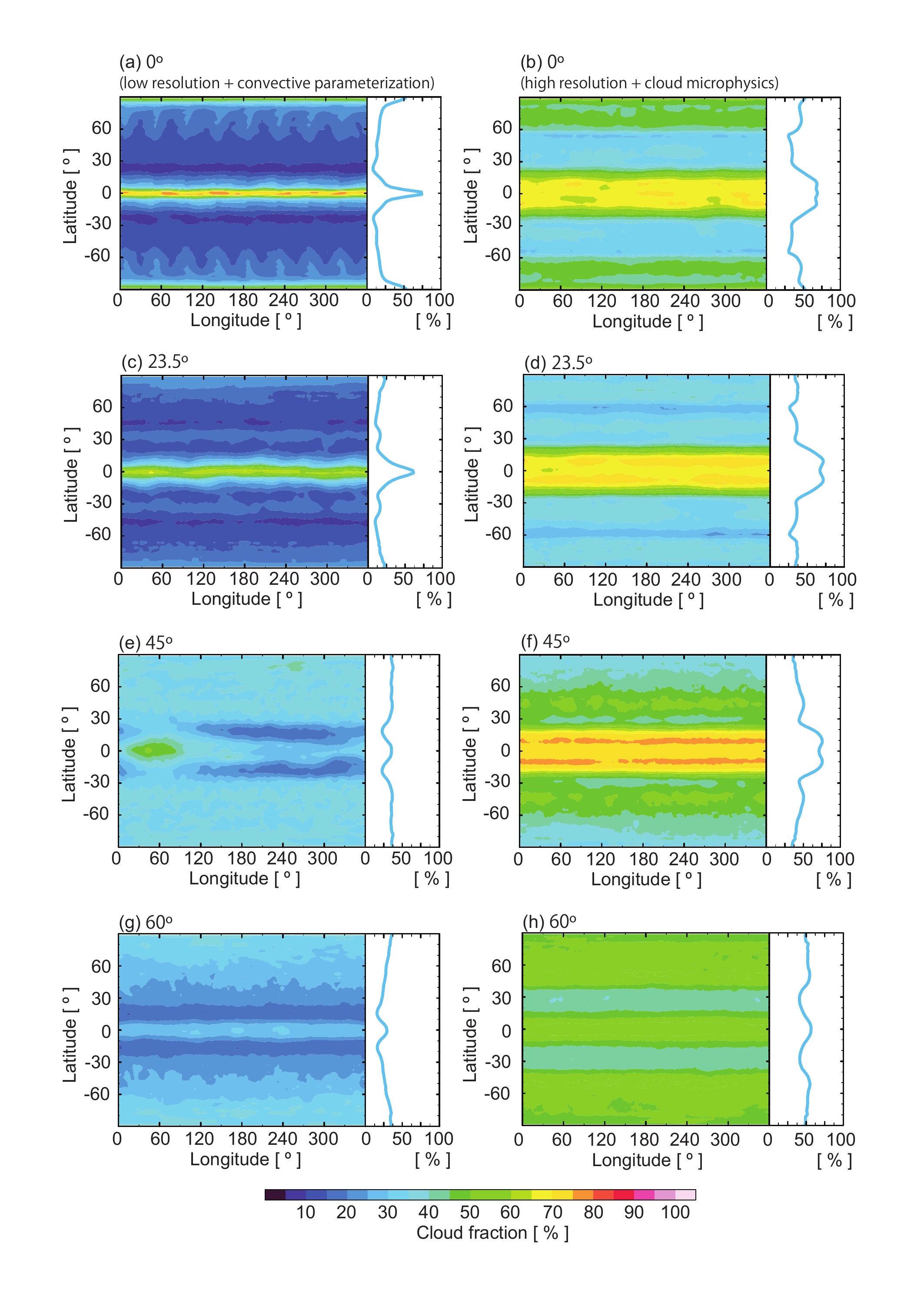}
\caption{Maps for cloud fraction of high clouds at low resolution with convective parameterization (left column) and at high resolution with explicit treatment of cloud microphysics (NSW6) (right column). Right panels beside maps show zonally averaged cloud fraction.}
\label{obl_fig9}
\end{center}
\end{figure*}

The difference in the surface temperature between low and high resolutions is strongly related to that in cloud contributions. Figure \ref{obl_fig7} shows the zonally averaged cloud radiative forcing, and Figures \ref{obl_fig8} and \ref{obl_fig9} show maps for cloud fraction of low and high clouds. Low-resolution simulations have a larger net radiative cooling than high-resolution simulations, in general (Fig. \ref{obl_fig7}). It is because that low-resolution simulations have stronger shortwave cooling and longwave warming than high-resolution simulations because low-resolution simulations have more low cloud fraction (Fig. \ref{obl_fig8}) and less high cloud fraction (Fig. \ref{obl_fig9}) but with its optical depth thicker (not shown).

Notably, for $\phi=45^\circ$ case with high resolution, the cloud shortwave radiative forcing has a convex shape due to less low cloud fraction (Fig. \ref{obl_fig8}f). It also has high cloud fraction in high clouds (Fig. \ref{obl_fig9}f), despite the weak contribution to radiative warming. These effects may be related to the warmer tropics in high-resolution simulations for $\phi=45^\circ$. In more detail, it is needed to analysis the formation of low and high clouds and the circulation, which will be addressed in a next task.

To further quantify the contributions of longwave cooling/warming from clouds and water vapor to the air temperature, we compute the effective surface emissivities associated with the clear sky and clouds and magnitude of the greenhouse effect (Table \ref{tab:emi}). The total effective surface emissivity ($\epsilon_\mathrm{all}$) is defined as the ratio of the outgoing longwave all sky radiation at the top of the atmosphere to that from the planetary surface. The magnitude of the greenhouse effect ($G$) is estimated as $G = 1 - \epsilon_\mathrm{all}$. The longwave effect of water vapor on the effective surface emissivity ($\epsilon_\mathrm{clear}$) can be described from the effect of clouds ($\epsilon_\mathrm{clouds}$) by $\epsilon_\mathrm{all} = \epsilon_\mathrm{clear} + \epsilon_\mathrm{clouds}$. The effect of clouds is defined as the ratio of the outgoing longwave clear sky radiation at the top of the atmosphere to the upward longwave radiation from the planetary surface \citep{Voigt&Marotzke2010}. 

The clear-sky components of the effective surface emissivities are smaller in high-resolution simulations for any obliquities, which means that the greenhouse effect due to larger amount of atmospheric water vapor contributes to more surface warming in high-resolution simulations. In fact, especially for low obliquity cases, this process can dominantly explain the larger greenhouse effect in high-resolution simulations. Meanwhile, for high obliquity cases, the contribution from clouds to the low effective surface emissivity gets larger in low-resolution simulations, which leads to the larger net greenhouse effect mainly by clouds in low-resolution simulations. To sum up, taken together with the shortwave radiative contributions, less low cloud fraction (i.e., lower planetary albedo) and more abundant water vapor make the surface warmer in high-resolution simulations.

\begin{longrotatetable}
\begin{deluxetable*}{ccccccccc}
\tablenum{1}
\tablecaption{Summary of global- and annual-mean climatic variables \label{tab:summary}}
\tablewidth{0pt}
\tablehead{
\colhead{Experiments} & \colhead{Surface} & \colhead{Precipitation} & \colhead{Water vapor} & \colhead{Low-level cloud} & \colhead{High-level cloud} & \colhead{Cloud radiative} & &  \\
& Temperature [K] & [mm/day] & amount [kg/$\mathrm{m}^2$] & fraction [\%] & fraction [\%] & forcing (long) [W/$\mathrm{m}^2$] &(short) [W/$\mathrm{m}^2$]& (net) [W/$\mathrm{m}^2$]}
\startdata
Low, $0^\circ$ & $271.07$ & $2.66$ & $10.14$ & $43.14$ & $19.53$ & $25.05$ & $-66.88$ & $-41.84$ \\
Low, $23.5^\circ$ & $275.14$ & $2.65$ & $10.37$ & $47.22$ & $20.18$ & $28.33$ & $-65.76$ & $-37.43$ \\
Low, $45^\circ$ & $293.68$ & $3.34$ & $24.21$ & $37.44$ & $32.01$ & $42.71$ & $-77.97$ & $-35.26$ \\
Low, $60^\circ$ & $290.72$ & $3.46$ & $22.66$ & $43.47$ & $24.13$ & $42.79$ & $-85.02$ & $-42.22$ \\
High, $0^\circ$ & $289.09$ & $3.91$ & $35.92$ & $23.05$ & $46.71$ & $21.13$ & $-54.06$ & $-32.93$ \\
High, $23.5^\circ$ & $291.34$ & $3.95$ & $34.08$ & $25.67$ & $46.28$ & $21.13$ & $-50.67$ & $-29.54$ \\
High, $45^\circ$ & $304.12$ & $4.77$ & $58.48$ & $14.45$ & $55.70$ & $17.35$ & $-33.59$ & $-16.24$ \\
High, $60^\circ$ & $303.57$ & $4.88$ & $69.99$ & $15.29$ & $48.91$ & $20.00$ & $-41.44$ & $-21.43$ \\
\enddata
\end{deluxetable*}
\end{longrotatetable}

\begin{deluxetable*}{ccccc}
\tablenum{2}
\tablecaption{The effective surface emissivities and magnitude of greenhouse effect\label{tab:emi}}
\tablewidth{0pt}
\tablehead{
\colhead{Experiments} & \colhead{Effective surface emissivity (total)} & \colhead{(clear sky)} & \colhead{(cloud)} & \colhead{Greenhouse effect} }
\startdata
Low, $0^\circ$ & $0.689$ & $0.755$ & $-0.0667$ & $0.311$ \\
Low, $23.5^\circ$ & $0.656$ & $0.730$ & $-0.0733$ & $0.344$ \\
Low, $45^\circ$ & $0.523$ & $0.628$ & $-0.105$ & $0.477$ \\
Low, $60^\circ$ & $0.529$ & $0.632$ & $-0.103$ & $0.471$ \\
High, $0^\circ$ & $0.643$ & $0.694$ & $-0.0512$ & $0.357$ \\
High, $23.5^\circ$ & $0.629$ & $0.680$ & $-0.0503$ & $0.371$ \\
High, $45^\circ$ & $0.565$ & $0.603$ & $-0.0379$ & $0.435$ \\
High, $60^\circ$ & $0.551$ & $0.590$ & $-0.0390$ & $0.449$ \\
\enddata
\end{deluxetable*}

Whether the climatological mean states, which are characterized by meridional overturning circulations, for example, become an ITCZ-like regime or a monsoon regime (i.e., one-cell meridional circulations with an off-equatorial upward branch; see Fig. 8 of \citet{Geen+2020}) would depend on the differences in meridional temperature gradient and/or the radiative feedback. Our results suggest that this regime separation even for the same insolation pattern may be affected by the differences in the energy distributions between conventional GCM and GCRM (global cloud-resolving model) modes, which is related to representation of clouds in middle- and high-latitudes. Further detailed analyses are needed to understand the mechanism of this issue in the future.

\section{Discussion} \label{sec:discussion}

\subsection{$\mathrm{CO}_{2}$ content and continents}

In our simulations, we assumed an Earth-like atmospheric composition. However, in Earth's history, the atmospheric composition has varied over the geological timescale. The atmospheric composition is closely related to how a planet accumulates and acquires volatiles and the subsequent atmospheric evolution. \cite{william&pollard2003} investigated the global mean surface temperature for high obliquity planets with a $\mathrm{CO}_{2}$ concentration 10 times higher than the present level. As expected, the global mean surface temperature increases with rising $\mathrm{CO}_{2}$ concentration. This trend would appear if we assume a higher $\mathrm{CO}_{2}$ concentration in our simulations. They showed a trend of decreasing global mean surface temperature with increased obliquity in cases with a high $\mathrm{CO}_2$ concentration because of the presence of ice and snow. In cases with a higher $\mathrm{CO}_{2}$, they also found a smaller seasonal amplitude because of prevention of low insolated area from cooling. Our results suggest that a simulation with a high resolution and an explicit treatment of cloud microphysics has more water vapor content than that with a low resolution and convective parameterization. With a higher $\mathrm{CO}_{2}$ concentration, whether such a large amount of water vapor plays a role as a greenhouse gas or makes the planetary albedo higher via cloud formation is still unclear. If the contribution of water vapor as a greenhouse gas is larger for a higher $\mathrm{CO}_{2}$ concentration, the amplitude of a seasonal cycle would be smaller than that for a low $\mathrm{CO}_{2}$ concentration.

The location and size of continents are also important for the climate. The continental temperature responds rapidly to insolation changes during the seasonal cycle. When we assume a continent at the tropics, the amplitude of the seasonal cycle is smaller than that with a continent at high latitudes because the change in insolation at low latitudes is smaller than that at high latitudes. Our simulation assumed an aqua planet configuration. When continents are added to our simulations, the hydrological cycle should change because of changes in the heat distribution as a heat source of the atmosphere. If the gradient of the surface temperature is attributed to the formation of the climatic regime, differences in thermal inertia become important. 

\subsection{Exoplanets}
As described in the section \ref{sec:intro}, terrestrial exoplanets within the habitable zone around M-type stars are thought to be tidally-locked rotators. These planets have permanent day-side and night-side hemispheres. Recently, climates for tidally locked terrestrial planet have been investigated using GCMs, which showed that their climates strongly depend on the planetary rotation period \citep{Haqq-Misra+2018, Kopparapu+2016, Kopparapu+2017}. A slow rotator has a mean zonal circulation from the day side to the night side, so-called the stellar-anti stellar circulation, with a thick cloud deck around the sub-stellar point. A rapid rotator has a weak convective motion and banded cloud formation around the sub-stellar point. Between them, the regime is called the Rhines rotation regime and it has a thermally direct circulation from the day-side to night-side with turbulence-driven zonal jets in the middle latitudes.

The amount of water vapor in the atmosphere is also important for the inner edge of the habitable zone. A planet with a wet atmosphere has a limit of the planetary thermal outgoing radiation, called the Simpson-Nakajima limit, caused by a wet atmospheric structure \citep{Ingersoll1969, Nakajima+1992}.

Therefore, the distributions of clouds and water vapor have significant roles in the climate of potentially habitable exo-terrestrial planets. Recently, the climates of such planets have been investigated using a high-resolution regional and global model with a convective parameterization scheme \citep{Wei+2020, Sergeev+2020}. \cite{Sergeev+2020} suggested that a model with convective parameterization may overestimate the heat re-distribution efficiency between hemispheres. Our results show that warmer climates in a high resolution with an explicit cloud treatment than in a low resolution with parameterizations for clouds. To investigate such potentially habitable exo-terrestrial planets in the near future, we need to use a global cloud resolving model to estimate the planetary albedo and the day-night temperatures, related to the distribution of clouds and water vapor on both hemispheres.

\section{Summary} \label{sec:summary}
Planetary climates are strongly affected by obliquity because it directly affects the seasonal change of insolation. Previous studies showed that a high-obliquity planet should have extreme seasonal cycles. Although climates for high obliquity planets have been investigated mainly using the energy-balance climate model (EBCM), a recent increase in computer resources enables us to address this issue with a three-dimensional general circulation model (GCM). Traditionally, a conventional GCM  with a $O(10^2)$-km horizontal mesh uses a cumulus parameterization and a large scale condensation scheme to evaluate cloud-related processes because it cannot explicitly resolve the coupling between clouds and dynamics.

In this study, we introduce a three-dimensional global non-hydrostatic model, named as NICAM (the Non-hydrostatic ICosahedral Atmospheric Model), which can explicitly compute the vertical moisture transport and cloud-system distributions. Using an aqua-planet configuration with a slab ocean model, we investigate the climatological mean states of temperature, precipitation, and large-scale circulations for planets with various obliquities ($0^{\circ}$, $23.5^{\circ}$, $45^{\circ}$, and $60^{\circ}$). Our simulations were conducted with two different horizontal resolutions: $1$) low resolution ($\sim 220$-km mesh) with  parameterization for clouds and $2$) high resolution ($\sim 14$-km mesh) with explicit treatment for cloud microphysics.

For low-resolution cases with parameterizations for clouds, the simulated  climatological states are in good agreement with those of previous studies. Planets with low obliquities have heavier precipitation around the equator than at middle and high latitudes. On the other hand, a high-obliquity planet has relatively strong precipitation at middle and high latitudes from the end of the summer to the winter. For high-resolution runs with the explicit treatment for cloud microphysics, the surface temperature is warmer than that in low resolution cases. A larger column water vapor content leads to heavier precipitation, although the precipitation pattern is similar to that in low resolution cases. For a $\phi = 45^{\circ}$ case, the meridional surface temperature gradient is inverted with respect to the low-resolution run. In a low-resolution case with parameterization for clouds, the case with $\phi = 45^{\circ}$ has one cell circulation, whereas that with high resolution has two cells circulation. The difference in the surface temperature between low and high resolutions is related to the cloud distribution and the amount of water vapor in the atmosphere. A low-resolution case with parameterizations for clouds generally has a larger net radiative cooling than a high-resolution case with explicit treatment for clouds microphysics, leading a warmer climate in high-resolution cases. This climatic difference should be caused by the energy redistribution in GCM and GCRM modes.

A caveat of this study is that the results of comparison between conventional GCM and GCRM modes can depend on tuning of the cumulus parameterization and explicit cloud scheme. Because how moisture is consumed and remained in the atmosphere is easily controlled by their tuning, the differences between GCM and GCRM modes that are presented here are one of possible solutions, rather than uniquely determined. Thus, we do not intend to emphasize the superiority of GCRM modes compared to GCM modes. Nevertheless, our results suggest that high-resolution simulations in which vertical moisture transportation and cloud formation are explicitly simulated may provide new physical interpretation that is largely different from that based on conventional GCM simulations.

Our study shows the impact of cloud-related processes on the climatological states due to differences in model resolutions and treatments for clouds. Although the reasons for the differences in the climatological states between the GCM and GCRM modes should be clarified in terms of the energy balance and transport, how cloud-related processes such as convection are treated is important for potentially habitable exoplanets.

\begin{acknowledgments}
We thank the editor and Dr. Jun Yang as a reviewer for their constructive comments and suggestions. This work was supported by MEXT KAKENHI Grants JP21K13975, JP19K03966, JP19H05703, and the Astrobiology Center of National Institutes of Natural Sciences (NINS) (Grant Number AB031014).
This study is supported by the Cooperative Research Activities of Collaborative Use of Computing Facility of the Atmosphere and Ocean Research Institute, the University of Tokyo and by the PPARC joint research program of Tohoku University. This work was supported by MEXT as “Program for Promoting Researches on the Supercomputer Fugaku” (JPMXP1020200305, Large Ensemble Atmospheric and Environmental Prediction for Disaster Prevention and Mitigation, ID:hp200128/hp210166) and used computational resources of supercomputer Fugaku provided by the RIKEN Center for Computational Science.
\end{acknowledgments}

%






\bibliography{ref_obl_nicam}{}
\bibliographystyle{aasjournal}



\end{document}